\documentclass[twocolumn]{aastex701}
\usepackage{amsmath}
\usepackage{color}
\def\beq{\begin{eqnarray}}
\def\eeq{\end{eqnarray}}

\begin{document}

\title{Hyperaccretion-driven relativistic jets from massive collapsars in active galactic nucleus disks}

\correspondingauthor{Tong Liu}
\email{tongliu@xmu.edu.cn}

\author[0000-0002-9130-2586]{Yun-Feng Wei}
\affiliation{Institute of Fundamental Physics and Quantum Technology, Ningbo University, Ningbo, Zhejiang 315211, China}
\affiliation{School of Physical Science and Technology, Ningbo University, Ningbo, Zhejiang 315211, China}
\email{weiyunfeng@nbu.edu.cn}

\author[0000-0001-8678-6291]{Tong Liu}
\affiliation{Department of Astronomy, Xiamen University, Xiamen, Fujian 361005, China}
\email{tongliu@xmu.edu.cn}

\author[0000-0002-4448-0849]{Bao-Quan Huang}
\affiliation{College of Intelligent Manufacturing, Nanning University, Nanning, Guangxi 530299, China}
\email{huangbaoquan@xmu.edu.cn}

\begin{abstract}
The observable characteristics of gamma-ray bursts (GRBs) embedded in the accretion disk of active galactic nuclei (AGNs) are mainly determined by the jet propagation within the disk. In the massive collapsar scenario, we consider that the mass and metallicity of progenitor stars can significantly affect the jet durations and luminosities, which in turn influence whether the jet can break out from AGN disks. For the cases with low metallicity, massive stars tend to keep their massive envelopes. Thus the hyperaccretion of these envelopes onto the newborn black holes (BHs) can prolong the activity duration of the central engine, thereby allowing the jets to potentially break out from the disks. For successful jets, we further study their prompt emission and afterglows for different supermassive BHs and locations and discuss the detectability of these signals by instruments such as \emph{Swift} and Einstein Probe. Future related observations will help constrain the structure, components, and evolutionary history of AGN disks and the massive stars embedded within them.
\end{abstract}


\section{Introduction}

Active galactic nuclei (AGNs) are thought to be powered by the accretion disks surrounding supermassive black holes \citep[SMBHs; e.g.,][]{LyndenBell1969,Shakura1973}. These disks are expected to be populated by numerous stars, either formed in the outer region of the disk via gravitational instability \citep[e.g.,][]{Paczynski1978,Kolykhalov1980,Goodman2003,Wang2011,Wang2012,Dittmann2020,Fan2023}, or captured from the nuclear star cluster  \citep[e.g.,][]{Artymowicz1993,Fabj2020,MacLeod2020,Wang2024}. Stars in AGN disks may evolve and produce compact objects such as black holes (BHs) and neutron stars. Due to the dense gas environment of the disk, massive stars and compact objects may undergo orbital migration and tend to accumulate in the inner regions of the disk \citep[e.g.,][]{McKernan2012,Bellovary2016}. Then, frequent core-collapse events of massive stars and binary mergers are likely to occur within the disk \citep[e.g.,][]{Cheng1999,Bartos2017,Stone2017,Yang2019,Tagawa2020,Li2021,Grishin2021,Ren2022,Li2023,Liu2024,She2025}. As a result, both long gamma-ray bursts (LGRBs) from the core-collapse of massive stars and short gamma-ray bursts (SGRBs) from compact binary mergers are expected to occur in the disk \citep[e.g.,][]{Perna2021,Zhu2021a,Zhu2021b,Zhang2024,Yuan2025}.

The observability of gamma-ray bursts (GRBs) occurring in AGN disks (hereafter AGN GRBs) has been investigated in many previous works \citep[e.g.,][]{Perna2021,Zhu2021b,Wang2022,Huang2024}. The emission of AGN GRBs is primarily determined by jet propagation within the AGN disk. Due to the high density and large-scale height of AGN disks, normal GRB jets are difficult to break out from the disk \citep[e.g.,][]{Zhu2021a,Zhu2021b,Huang2024}. Consequently, the photons from prompt emission and afterglow emission of AGN GRBs undergo diffusion and absorption before escaping the disk photosphere \citep[e.g.,][]{Perna2021,Wang2022,Ray2023,Kathirgamaraju2024}. For a choked jet, the jet energy would be stored in a hot cocoon, which can produce X-ray radiation when the cocoon breaks through the AGN disk \citep{Zhu2021b}. In addition, the forward shock driven by the choked jet can give rise to shock breakout emission. On the other hand, if the GRB jet succeeds in breaking out from the AGN disk, both ``delayed'' prompt and afterglow emission may be detected as normal \citep{Zhang2024}.

\citet{Perna2021} found that the GRB jets may be able to break out from the AGN disk when the SMBH has a relatively low mass and the GRB occurs close to it. In \citet{Yuan2025}, they consider the influence of wind from the GRB progenitor on GRB jet propagation in the AGN disk. The results show that the strong wind from the SGRB progenitors can penetrate the disk surface, and the jet can break out from the disk surface. However, previous studies have generally neglected the impact of the progenitor properties on jet characteristics, and instead employed typical GRB jet parameters to calculate the jet propagation in AGN disks. For massive progenitors, the jet duration can be strongly modified by progenitor factors such as mass and metallicity. In the collapsar scenario \citep[e.g.,][]{Woosley1993,Woosley2002,Woosley2012,Zhang2004,Liu2018,Song2019}, an observable GRB is triggered when the jet breaks out from the envelope of a massive star. Thus, the GRB duration is determined by both the lifetime of the central engine and jet propagation in the envelope. It is widely believed that Wolf-Rayet stars are plausible LGRB progenitors, since they have no hydrogen and helium envelopes, and thus jets can penetrate more easily \citep{Matzner2003}. Nevertheless, many studies suggest that metal-poor stars such as Pop III stars can also produce GRBs \citep[e.g.,][]{Suwa2011,Nagakura2012,Nakauchi2012,Nakauchi2013,Matsumoto2015}. Although they keep massive hydrogen envelopes until the pre-collapse phase, the accretion of these envelopes activates the central engine long enough for the jet to break out from the progenitor. \citet{Nakauchi2012} found that low-mass Pop III stars may produce ultra-long GRBs (ULGRBs) with durations around $10^{4}$ s. Correspondingly, variations in the density profiles of progenitor stars within AGN disks can result in significant differences in jet durations. In particular, ultra-long duration jets may break out from the AGN disk and give rise to GRB emission.

In this paper, we study the impact of progenitor properties on GRB jet propagation in AGN disks and predict the prompt and afterglow emissions from AGN GRBs. The paper is organized as follows. Section 2 introduces the models of the AGN disk, progenitor stars, hyperaccretion  processes, and jet propagation. We present the effects of progenitor mass and metallicity on observational characteristics of AGN GRBs and calculate their prompt and afterglow emissions in Section 3. Conclusions and discussion are made in Section 4.

\section{Model}

\subsection{AGN disks}

In this work, we adopt the standard AGN disk model described in \citet{Sirko2003}, hereafter the SG model, which considers the effect of self-gravity and provides an accurate description of the radial structure. We set the outer disk boundaries as $ R_{\rm{max}}=2\times 10^{5}\,R_{\rm g}$, where $R_{\rm{g}}=2GM_{\rm{SMBH}}/c^{2}$ is the Schwarzschild radius with $M_{\rm{SMBH}}$ being the mass of SMBHs, and $c$ being the speed of light. The masses of SMBHs are set to $M_{\rm{SMBH}}=10^6$, $10^7$, and $10^{8}\,M_{\odot}$. The luminosity of AGN disk radiation is set to $L_{0}=0.1~\dot{M}_{\rm{SMBH}}c^{2}$, where $\dot{M}_{\rm{SMBH}}$ is the mass accretion rate of the SMBH. Then the SMBH accretion efficiency is calculated as $l_{\rm{E}}=L_{0}/L_{\rm{Edd}}$, where $L_{\rm{Edd}}$ is the Eddington luminosity. The viscosity parameter is set to $\alpha =0.1$. Following the formulation of \citet{Sirko2003}, we solved the disk equations to determine the radial structure and physical properties of the disks, thereby obtaining the mid-plane density and disk scale height.

The density distribution of the AGN disk in the vertical direction can be approximated by an isothermal atmosphere \citep{Netzer2013},
\beq
\rho _{\rm{disk}}(R,R_{\rm z})=\rho _{\rm disk,0}(R) \exp [-\frac{R_{\rm z}^{2}}{2H(R)^{2}}],
\eeq
where $R$ is the radial radius, $R_{\rm z}$ represents the vertical distance from the equatorial plane, $\rho _{\rm disk,0}(R)$ is the mid-plane density, $H(R)$ is the half-thickness of
accretion disks.

\subsection{Progenitors and relativistic jets}

\begin{figure}
\centering
\includegraphics[angle=0,scale=0.3]{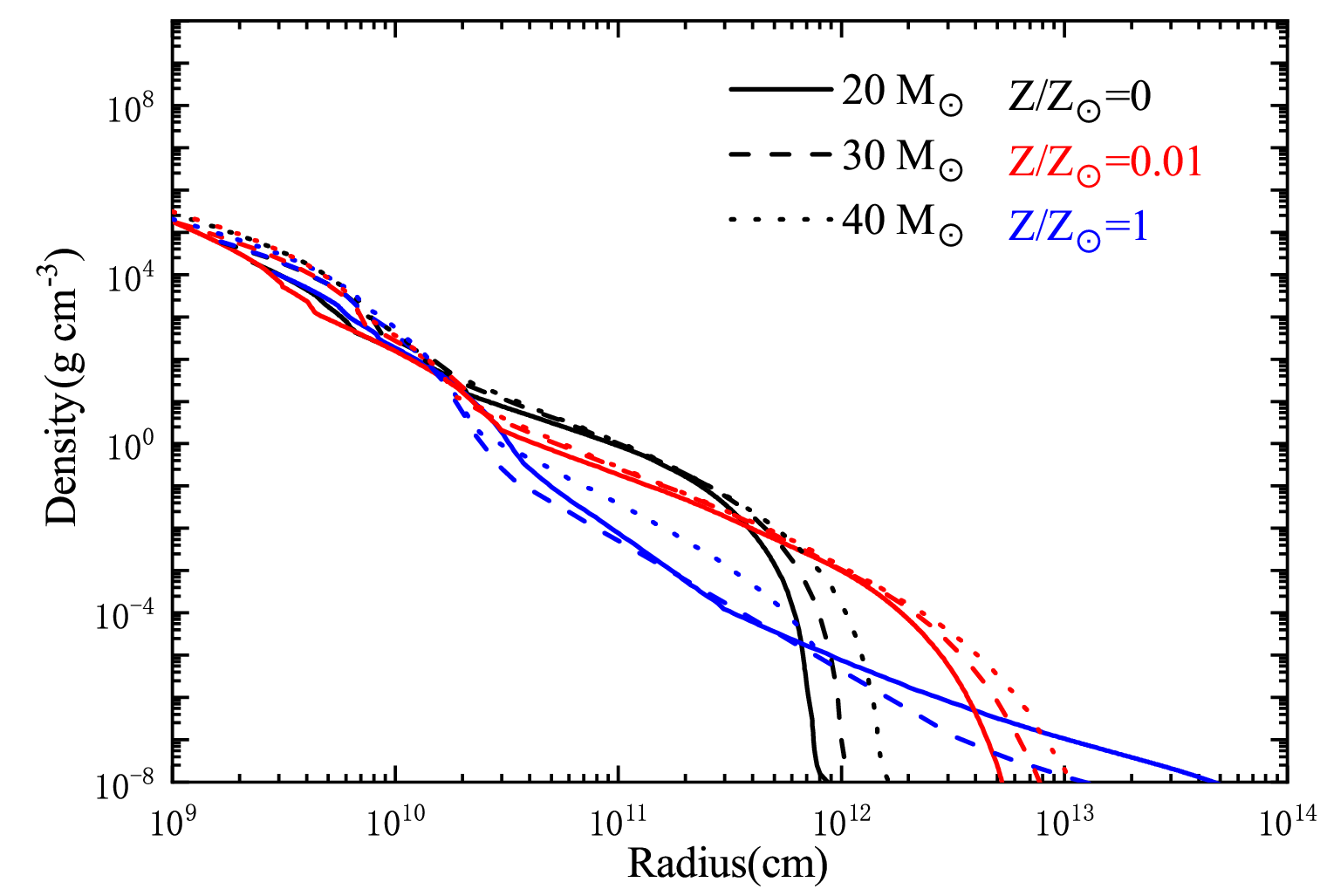}
\caption{Density profiles of the progenitor stars investigated, with different masses and metallicities. The black, red, and blue curves correspond to progenitor metallicities of $Z/Z_{\odot}=0, ~0.01$, and $1$, respectively. The solid, dashed, and dotted lines represent the progenitor masses of $M_{\rm{pro}}/M_{\odot}=20, ~30$, and $40$, respectively.}
\label{fig1}
\end{figure}

We employ the well-known presupernova (pre-SN) models with initial mass in the range of $20-40\,M_{\odot}$ as progenitor models \citep[e.g.,][]{Woosley2002,Woosley2007,Heger2010}. For those models, the models with zero metallicity ($Z/Z_{\odot} = 0$) and solar metallicity ($Z/Z_{\odot} = 1$) are referenced from \citet{Heger2010} and \citet{Woosley2007}, respectively, and the ones with metallicity $Z/Z_{\odot} =0.01$ are provided by Prof. Alexander Heger in private communication, where $Z_\odot$ represents the metallicity of the Sun. These models were evolved using the KEPLER code \citep{Weaver1978,Woosley2002} through all stable nuclear burning stages until their iron cores became unstable to collapse. The density profiles of the investigated models are shown in Figure \ref{fig1}. Black, red, and blue curves correspond to progenitor star metallicities of $Z/Z_{\odot}=0, ~0.01$, and $1$, respectively, while solid, dashed, and dotted lines represent the progenitor masses of $M_{\rm{pro}}/M_{\odot}=20, ~30$, and $40$, respectively. One can see that the density profile of the envelope is mainly determined by metallicity. Progenitors with solar metallicity lose their dominant masses before collapsing. This is due to the strong stellar wind that blows out of the hydrogen envelope, which may not be present in zero- or low-metal stars due to the low opacity in their envelopes \citep{Kashiyama2013}.

Using these density profiles, we can calculate the mass accretion rate of the progenitors. The accretion timescale of matter at a radius $r$ is estimated by the free-fall timescale \citep[e.g.,][]{Woosley2012,Matsumoto2015}. For each mass shell at radius $r$, the free-fall timescale can be calculated as
\beq
t_{\rm{ff}}=\sqrt{\frac{3\pi}{32G\bar{\rho}}}=\frac{\pi}{2}\sqrt{\frac{r^{3}}{2GM_{\rm{r}}}},
\eeq
where $\bar{\rho}=3M_{\rm{r}}/(4\pi r^{3})$ is the mean density within $r$, $M_{\rm{r}}$ is the mass coordinate. Then the mass accretion  rate is calculated by \citep[e.g.,][]{Suwa2011}:
\beq
\dot{M}=\frac{dM_{\rm{r}}}{dt_{\rm{ff}}}=\frac{dM_{\rm{r}}/dr}{dt_{\rm{ff}}/dt}=\frac{2M_{\rm{r}}}{t_{\rm{ff}}}(\frac{\rho}{\bar{\rho}-\rho }),
\eeq
where $\rho$ is the mass density of the progenitor star.

Relativistic jets are launched from a BH hyperaccretion system at the center of a massive collapsar. Two well-known mechanisms have been proposed for jet production: the neutrino-antineutrino annihilation process \citep[e.g.,][]{Popham1999,Narayan2001,Chen2007,Liu2017}, and the Blandford-Znajek (BZ; \citet{Blandford1977}) mechanism. For the same BH spin parameter and accretion rate, the BZ luminosity is roughly two orders of magnitude higher than the neutrino annihilation luminosity \citep{Liu2015,Liu2017}. Considering the large height of the AGN disk, it is difficult for the jet driven by the neutrino annihilation mechanism to break out from the AGN disk. Accordingly, we adopt the jet model based on the BZ process, in which the jet luminosity can be modeled following \citep[e.g.,][]{Komissarov2010,Suwa2011,Matsumoto2015}
\beq
L_{j}=\eta \dot{M}c^{2},
\eeq
where $\eta$ is the efficiency parameter. For simplicity, we assume that $\eta$ is constant and take its value of $\eta=6.2\times 10^{-4}$ \citep{Suwa2011}.

It should be noted that the pre-SN models employed here were developed for the typical density of interstellar medium (ISM), rather than for the much higher densities of AGN disks. Actually, the AGN disk and the stars therein are interdependent and interactive. The formation, evolution, and death of stars in AGN disks are expected to differ from those in normal environments.

\subsection{Jet propagation}

\begin{figure*}
\centering
\includegraphics[angle=0,scale=0.6]{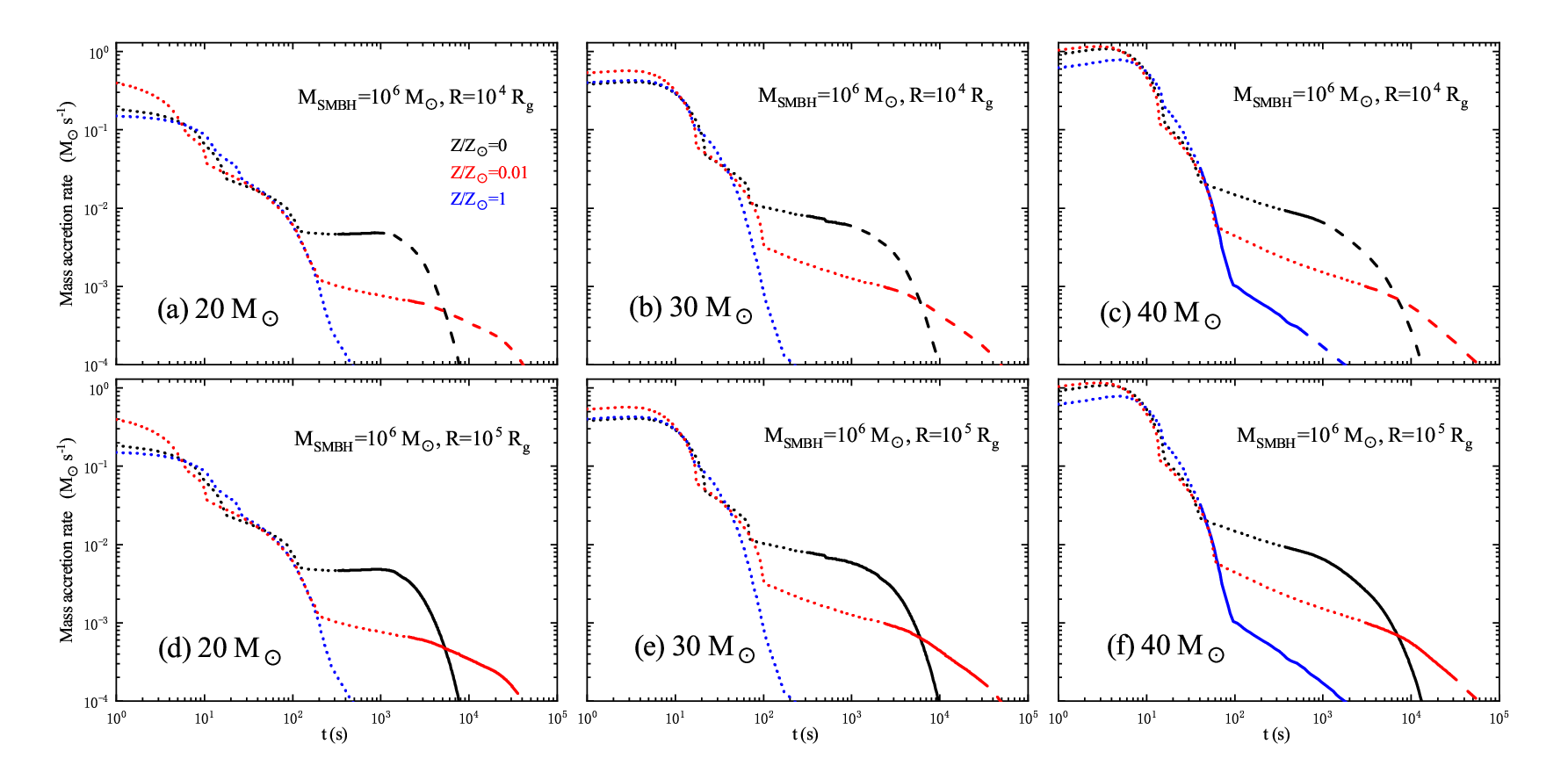}
\caption{Time evolution of the mass accretion rate onto the newborn BH. The black, red, and blue curves correspond to the cases of progenitors with metallicities of $Z/Z_{\odot}=0$, $0.01$, and $1$, respectively. For each color, the dotted lines represent the jet propagation inside the star, the solid lines indicate the jet propagation inside the AGN disk, and the dashed lines indicate the phase after the jet breaks out from the AGN disk. The dashed lines give the information of the duration and energetics of AGN GRBs. The left, middle, and right panels correspond to progenitor masses of $M_{\rm{pro}}/M_{\odot}$ = 20, 30, and 40, respectively, with an SMBH mass of $10^{6} M_{\odot}$. In top panels and bottom panels, the radial locations from the SMBH are $10^{4} R_{g}$ and $10^{5} R_{g}$, respectively.}
\label{fig2}
\end{figure*}

The relativistic jet launched by the central engine first propagates through the stellar envelope. After breaking out from the envelope, the jet enters the AGN disk and continues its propagation in a similar mode. The jet dynamics within dense medium has been widely investigated in previous works \citep[e.g.,][]{Matzner2003,Bromberg2011,Nakauchi2012,Kashiyama2013,Nagakura2014,Matsumoto2015,Liu2018,Yu2020,Wei2022}. A relativistic jet launched by the central engine collides with the dense medium and forms forward and reverse shocks at its head. The velocity of the jet head can be estimated from the pressure balance at the interface between the jet and the dense medium as \citep{Matzner2003,Bromberg2011}
\beq
\beta _{h}=\frac{1}{1+\tilde{L}^{-1/2}},
\eeq
where
\beq
\tilde{L}\equiv \frac{L_{j}(t-r_{{h}}/c)}{\pi r_{{h}}^{2}\theta _{j}^{2}\rho (r_{{h}})c^{3}},
\eeq
and $\theta _{j}$ is the jet half-opening angle. Here, we assume that the half-opening angle of the jet is constant at $\theta _{j}=5^{\circ}$. Then, the position of the jet head is given by $r_{\rm h}(t)=\int c\beta_{\rm{h}} dt$.

If the velocity of the jet head is nonrelativistic, $ \beta _{h}<0.3$, a significant fraction of the shocked material will spread out sideways to form a cocoon around the jet \citep{Bromberg2011}. The pressure in the cocoon is sustained by a continuous flow of energy from the jet head. The energy stored in the cocoon is given by
\beq
E_{c}(t)=\eta _{c}\int_{t_{in}}^{t-r_{h}(t)/c}L_{j}(t^{'})dt^{'},
\eeq
where $\eta _{c}$, varies between $0$ and $1$, denotes the fraction of the energy that flows into the cocoon, $t_{in}$  represents the jet launching time, corresponding to the initial position $r_{in}$. During the propagation of the jet within the stellar envelope, one can reasonably assume that $\eta _{c}\sim1$, since the velocity of the jet head is nonrelativistic for the majority of the time. Considering the pressure balance at the surface of the cocoon, the expansion velocity of the cocoon is given by
\beq
\beta _{c}=\sqrt{\frac{P_{c}}{\rho _{*}c^{2}}},
\eeq
where $P_{c}$ is the pressure of the cocoon, $\rho _{*}=M_{r_{h}}/(4\pi r_{h}^{3}/3)$ is the mean density of the progenitor star in the scale of the jet head. Since the cocoon pressure is radiation-pressure dominated, it can be expressed as $P_{c}=E_{c}/3V_{c}$, where $V_{c}$ represents the volume of the cocoon. We assume that the shape of the cocoon is a cone with the height of $r_{h}(t)$ and the base radius of $r_{c}(t)$, so that $V_{c}(t)=\pi r_{c}^{2}(t)r_{h}(t)/3$. Substituting the above expressions into Equation (8), the expansion velocity of the cocoon can be evaluated as
\beq
\beta _{c}=\frac{r_{h}(t)}{r_{c}(t)}\sqrt{\frac{4\eta _{c}\int_{t_{in}}^{t-r_{h}(t)/c}L_{j}(t^{'})dt^{'} }{3(M_{r_{h}}-M_{r_{in}})}}.
\eeq
The position of the cocoon edge is calculated as $r_{c}(t)=\int_{0}^{t}\beta _{c}(t^{'})c\,dt^{'}$.

For those successful collapsar jets, the jet head accelerates rapidly after entering the hydrogen envelope, and the velocity of the jet head is always larger than that of the cocoon surface except during the very early time \citep{Meszaros2001,Nakauchi2012,Matsumoto2015}. Then, the jet head propagates into the AGN disk with a relativistic velocity. Since the density of the disk medium is far lower than that of the stellar envelope, the effect of the cocoon on the jet is weak \citep{Bromberg2011}. If the jet head arrives at the surface of the AGN disk before the central engine ceases, the jet can successfully break out from the disk.

\begin{figure*}
\centering
\includegraphics[angle=0,scale=0.6]{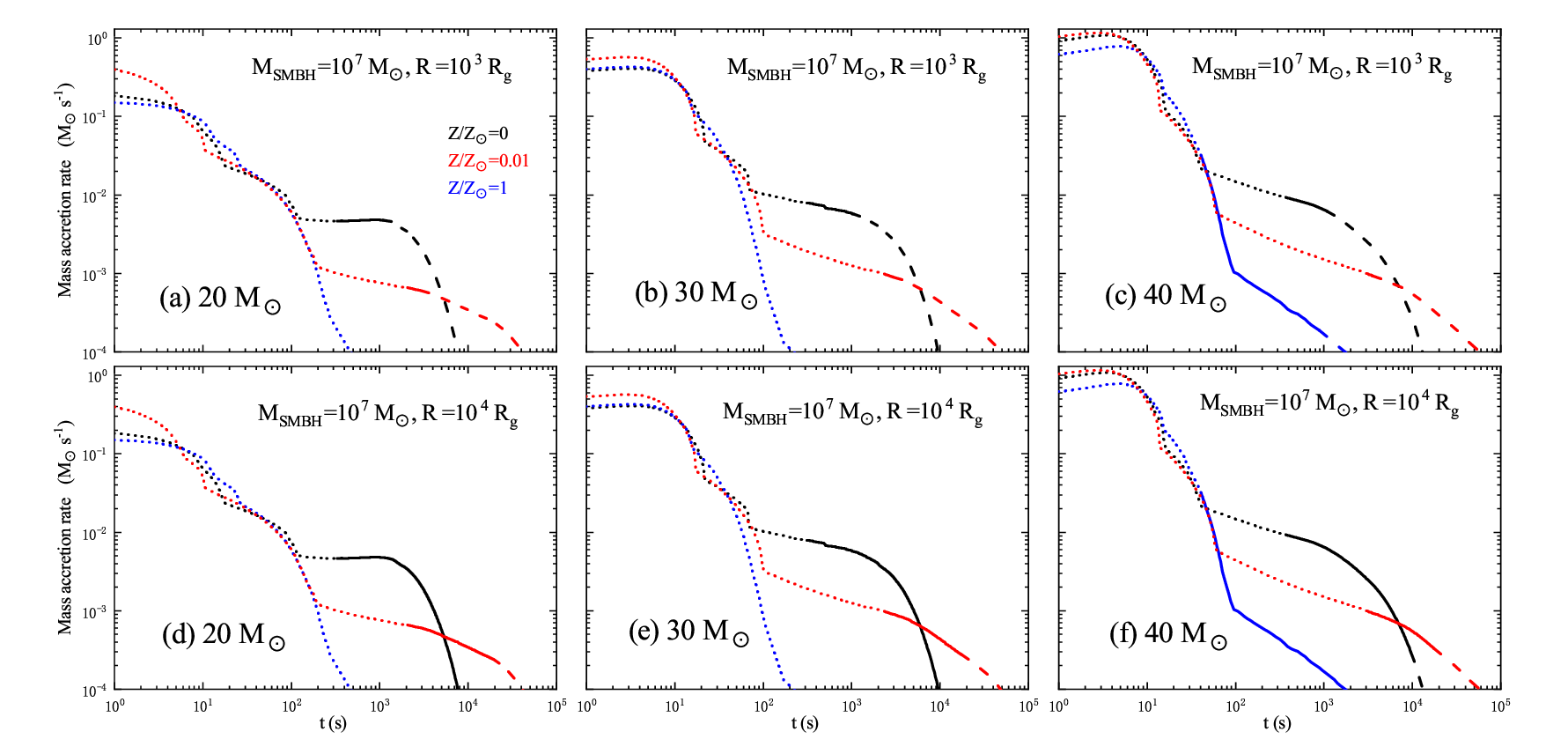}
\caption{Same as Figure \ref{fig2}, but for an SMBH mass of $10^{7} M_{\odot}$. The top and bottom panels correspond to radial locations of $10^{3} R_{g}$ and $10^{4} R_{g}$ from the SMBH, respectively.}
\label{fig3}
\end{figure*}

\begin{figure*}
\centering
\includegraphics[angle=0,scale=0.29]{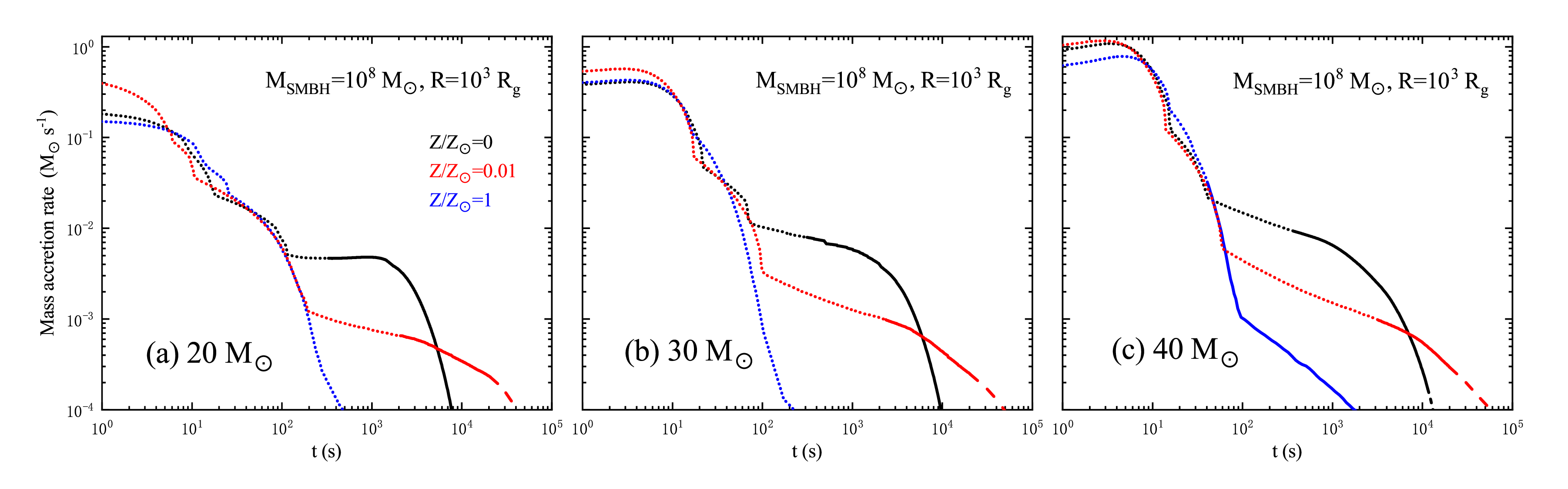}
\caption{Same as Figure \ref{fig2}, but for an SMBH mass of $10^{8} M_{\odot}$. The radial location from the SMBH is $10^{3} R_{g}$.}
\label{fig4}
\end{figure*}

In Figure \ref{fig2}, we display the mass accretion rate onto the newborn BH, which characterizes the evolution of the central engine. The horizontal axis is the time from jet formation. The black, red, and blue curves correspond to progenitors with metallicity of $Z/Z_{\odot}=0$, $0.01$, and $1$, respectively. For each color, the dotted lines represent the jet propagating inside the star, the solid lines represent the jet propagating within the AGN disk, and the dashed lines correspond to the time after the jet breaks out of the AGN disk. Dashed lines give information of the duration and energy of AGN GRBs. The left, middle, and right panels correspond to progenitor star masses of $M_{\rm{pro}}/M_{\odot}$= 20, 30, and 40, respectively. The SMBH mass is $10^{6} M_{\odot}$. In the top panels and bottom panels, the radial distances from the SMBH is $10^{4} R_{g}$ and $10^{5} R_{g}$, respectively. In Figures \ref{fig3} and \ref{fig4}, we show the results for SMBH masses of $10^{7} M_{\odot}$ and $10^{8} M_{\odot}$, respectively. It can be seen that the mass accretion rate is primarily determined by the density profile of the progenitor star.

The progenitor mass and metallicity shape the density profiles of the progenitors, which in turn affect the activity duration of the central engine and ultimately determine whether the jet can successfully break out from the AGN disk. As shown in Figure \ref{fig2}, for progenitors with a metallicity of $Z/Z_{\odot}=0.01$, their central engines tend to have relatively longer activity durations, and the jets are more likely to break out from the AGN disk. In contrast, for metallicity $Z/Z_\odot = 0$, the central engines exhibit shorter durations but higher luminosities. Progenitors with solar metallicity are more likely to produce failed jets due to the relatively short lifetime of the central engine. The effect of progenitor mass on the evolution of the accretion rate is relatively weak. Here, we consider progenitor masses up to $40\,M_{\odot}$ and do not further include more massive progenitors, as such stars are expected to have larger radii, resulting in longer jet breakout times and weaker luminosities at breakout \citep{Nakauchi2012}. Consistent with previous studies, when the SMBH mass is relatively low ($M_{\rm{SMBH}} \sim 10^{6} M_{\odot} $) and the GRB occurs closer to the SMBH, the jet is more likely to break out of the disk. Figure \ref{fig2} shows that for $M_{\rm{SMBH}}\sim 10^{6} M_{\odot} $, the jet breakout region can reach as far as $10^{5} R_{g}$. In Figure \ref{fig4}, all jets fail to break out from the disk at $10^{4} R_{g}$, so only the cases at $10^{3} R_{g}$ are presented.

As mentioned above, the mass and metallicity of the progenitor star play a key role in determining whether the jet can break out of the AGN disk. It is therefore interesting to discuss the fraction of GRB jets that are able to break out from the AGN disk. At a given radial position, the progenitor metallicity sets the minimum progenitor mass $M_{\rm min}$ required for the jet to break out of the AGN disk. Here, we consider two extreme cases of metallicity, $Z/Z_\odot = 0$ and $Z/Z_\odot = 1$. For $M_{\rm{SMBH}}=10^{7} M_{\odot} $ and $R=10^{3} R_{g}$, $M_{\rm min}\sim20 M_{\odot}$ and $\sim40 M_{\odot}$ for $Z/Z_\odot = 0$ and $Z/Z_\odot = 1$, respectively. The maximum progenitor mass $M_{\rm max}$ permitting jet breakout is approximately $45 M_{\odot}$, as jets launched from more massive progenitor stars tend to have lower luminosities. The mass range of progenitor stars that can produce GRB jets also depends on the metallicity. Here we adopt a progenitor mass range of $20-60 M_{\odot}$ for zero metallicity, and $40-60 M_{\odot}$ for solar metallicity \citep{Nakauchi2012}. Considering the initial mass function (IMF) as $ \xi(M)\propto M^{-\Gamma }$, where $\Gamma\sim 1.0$ is the index of the ``top-heavy'' IMF for AGN disks \citep{Toyouchi2022}, we roughly estimate the fraction of successful jets. The results show that approximately 73.9 \% of the jets can break out of the AGN disk for zero metallicity, whereas about 29.6 \% can do so for solar metallicity. Of course, since not all GRB jets are oriented perpendicular to the disk plane, the actual fraction is expected to be lower. A more accurate prediction of the fraction of successful jets would require propagation calculations for a larger sample of massive collapsars, which represents an interesting direction for future work.

\section{AGN GRBs}

Here, we mainly focus on the GRB emission produced by jets that successfully break out of the AGN disk; in other words, the central engine is still active after the jet breakout. Accordingly, we assume that both the prompt and afterglow emissions are produced outside the AGN disk. The density structure above the AGN disk remains uncertain. For simplicity, we assume that the density in the region outside the disk is approximately that of ISM, and we calculate the GRB emission accordingly. The emission from jets that are choked within the AGN disk is not included in this work. In such cases, where the central engine has a relatively short duration, the prompt and afterglow emissions are expected to be produced inside the AGN disk \citep{Perna2021,Lazzati2022}. Then, the intrinsic prompt and afterglow emission are affected by local disk properties. Moreover, before escaping from the AGN disk, the radiation will undergo scattering and absorption, which has been statistically investigated in \citet{Kang2025}. If diffusion is not dominant, the radiation is expected to be dominant in prompt gamma-ray emission. When diffusion becomes significant, the radiation would more likely be detected as afterglow emission in X-ray band. Specifically, the observable signals highly depend on the AGN disk models, the SMBH mass, and the locations.

\subsection{Prompt emission}

We assume that the relativistic jet can contribute to the prompt high-energy emission once it breaks out from the AGN disk. In the rest frame, the duration of GRB is estimated as $ t_{\rm{GRB}}=t_{\rm{eng}}-t_{\rm{b}}$, where $t_{\rm{eng}}$ denotes the duration of the central engine and $t_{\rm{b}}$ denotes the jet breakout time. Following \citet{Nakauchi2012}, we set a $10\%$ efficiency for converting the jet energy into radiation. The isotropic radiated energy of the prompt emission, $E_{\rm\gamma,iso}$, and the peak luminosity, $L_{\rm p}$, are then given by
\beq
E_{\rm\gamma,iso}=\frac{2}{\theta _{\rm j}^{2}}\int_{t_{\rm b}}^{t_{\rm eng}}\zeta L_{\rm j}(t)dt,
\eeq
and
\beq
L_{\rm p}=L_{\rm \gamma,iso}(t_{\rm b})=\frac{2}{\theta _{\rm j}^{2}} L_{\rm j}(t_{\rm b}),
\eeq
where $\zeta=0.1$.

The luminosity is expected to peak at breakout, as the mass accretion rate decreases monotonically afterward. We then evaluate the spectral peak energy of the prompt emission in the observer frame, $E_{\rm p}^{\rm obs}$. We adopt two typical empirical correlations of LGRBs: the $E_{\rm p}-L_{\rm p}$ correlation \citep{Yonetoku2004} and the $E_{\rm p}-E_{\rm \gamma,iso}$ correlation \citep{Amati2002}. These two correlations can be expressed as
\beq
\frac{L_{\rm p}}{10^{52}\,\rm erg\,s^{-1}}\sim 2\times 10^{-5}\left [ \frac{E_{\rm p}^{\rm obs}(1+z)}{1\, \rm keV} \right ]^{2},
\eeq
and
\beq
\left [ \frac{E_{\rm p}^{\rm obs}(1+z)}{1\, \rm keV} \right ]\sim80\left [ \frac{E_{\rm \gamma,iso}}{10^{52}\,\rm erg} \right ]^{0.57},
\eeq
respectively. Here, we set the AGN redshift $z = 1$.

Then we discuss the detectability of AGN GRBs with instruments such as the \emph{Swift} Burst Alert Telescope (BAT) \citep{Barthelmy2005} and Einstein Probe (EP) \citep{Yuan2025}. The BAT covers the energy range from $15-150$ keV. The EP carries two instruments, the Wide-field X-ray Telescope (WXT; $0.5-4$ keV \citep{Yuan2022}) and the Follow-up X-ray Telescope (FXT; $0.3-10$ keV \citep{Chen2021}). The energy flux detected by the detector is given by
\beq
f_{\rm sig}(t_{\rm obs})=\frac{L_{\rm \gamma,iso}(t_{\gamma })}{4\pi d_{\rm L}(z)^{2}}\frac{\int_{E_{\rm min}}^{E_{\rm max}}EN(E)dE}{\int_{E_{0}}^{E_{\infty }}EN(E)dE},
\eeq
where $L_{\rm \gamma,iso}(t_{\gamma })$ denotes the isotropic equivalent luminosity of the burst at the time from the breakout,  $t_{\rm obs}=(1+z)t_{\gamma}$ is the time in the observer frame, $N(E)$ is the Band spectrum \citep{Band1993} with the spectral indices of $ \alpha =-1$ and $\beta =-2.3$ \citep{Kaneko2006}, and $d_{\rm L}(z)$ is the luminosity distance.

\begin{figure}
\centering
\includegraphics[angle=0,scale=0.32]{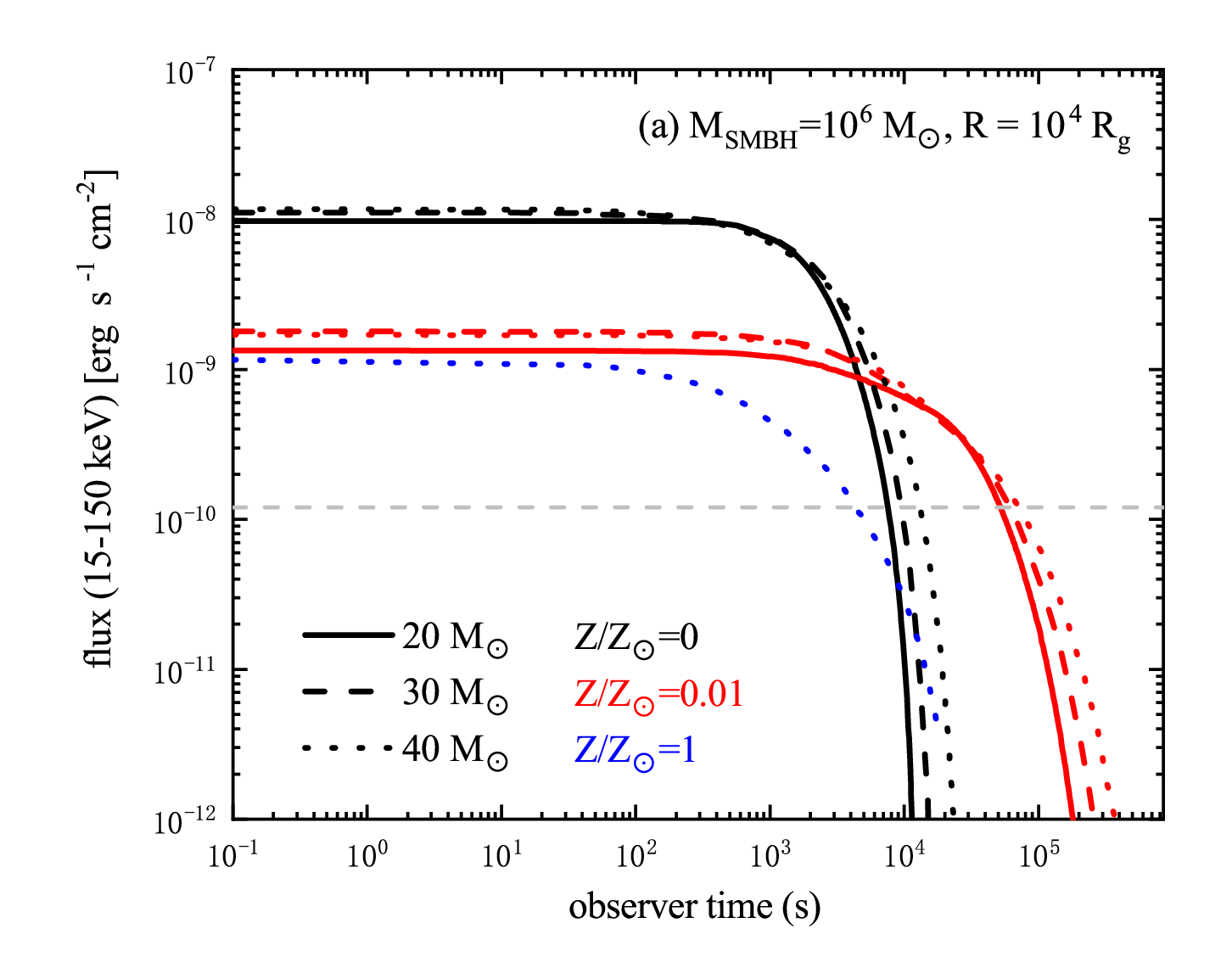}
\includegraphics[angle=0,scale=0.32]{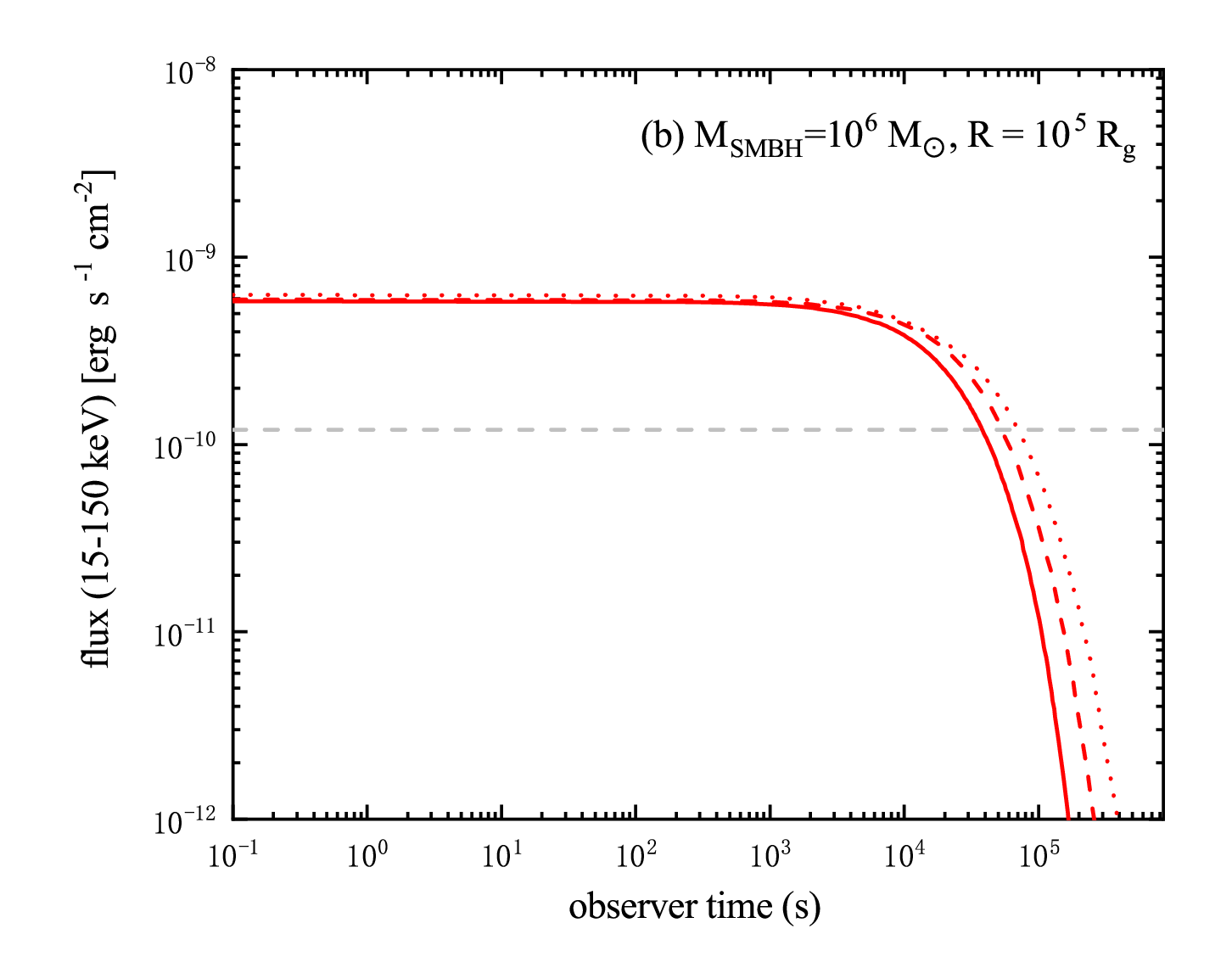}
\caption{Light curves of the prompt emission of AGN GRBs with different progenitors. The flux is calculated at the Swift BAT energy range ($15-150$ keV). The black, red, and blue curves correspond to the cases of progenitors with metallicity of $Z/Z_{\odot}=0$, $0.01$, and $1$, respectively. The solid, dashed, and dotted lines represent the progenitor masses of $M_{\rm{pro}}/M_{\odot}$=20, 30, and 40, respectively. The gray dotted lines indicates the BAT sensitivities $f_{\rm sen} (\Delta t_{\rm obs})$ with the integration times of $\Delta t_{\rm obs}=10^{4}$ s. The SMBH mass is $10^{6} M_{\odot}$. In panels (a) and (b), the radial locations from the SMBH are $10^{4} R_{g}$ and $10^{5} R_{g}$, respectively.}
\label{fig5}
\end{figure}

\begin{figure}
\centering
\includegraphics[angle=0,scale=0.32]{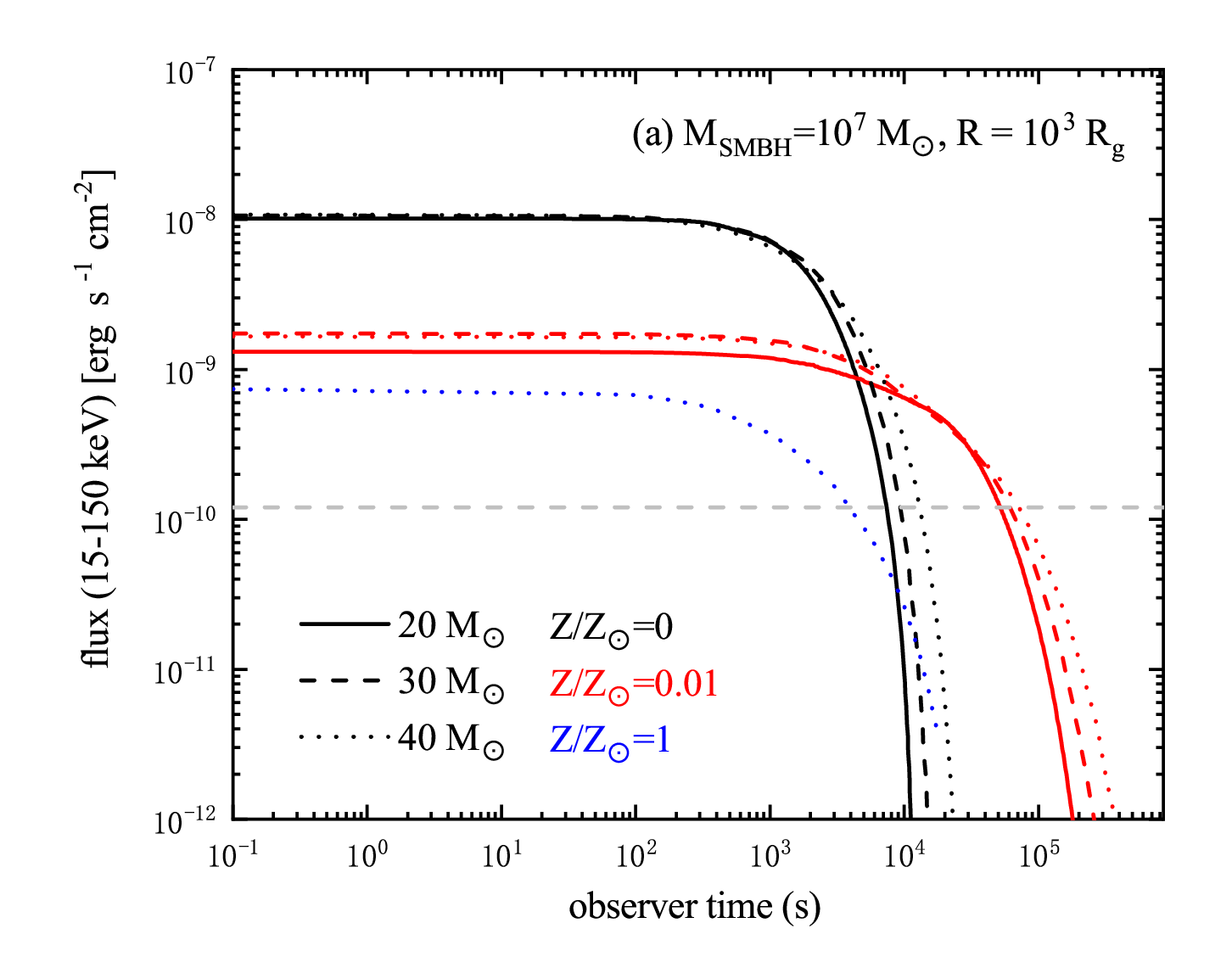}
\includegraphics[angle=0,scale=0.32]{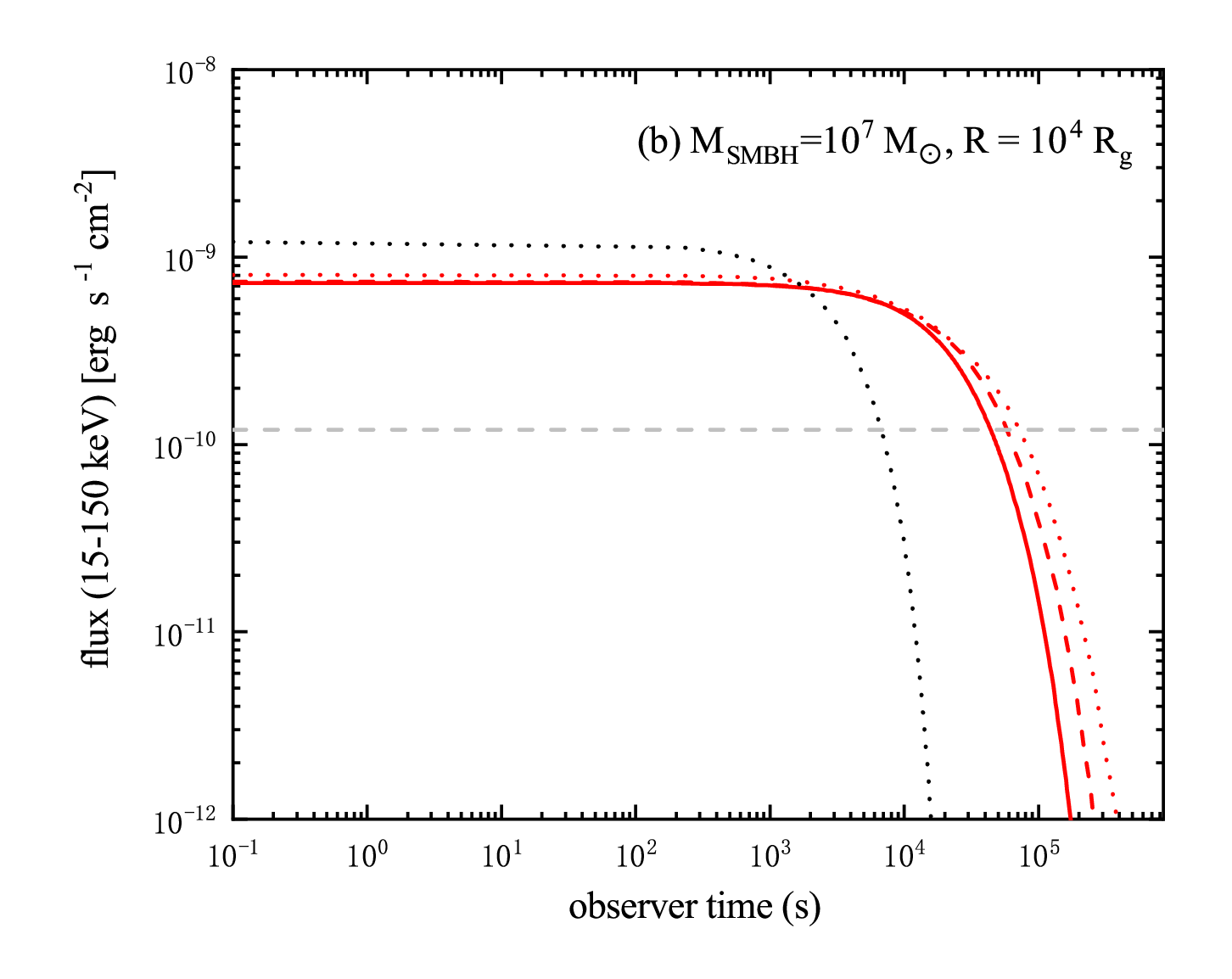}
\caption{Same as Figure \ref{fig5}, but for an SMBH mass of $10^{7} M_{\odot}$. In panels (a) and (b), the radial locations from the SMBH are $10^{3} R_{g}$ and $10^{4} R_{g}$, respectively.}
\label{fig6}
\end{figure}

\begin{figure}
\centering
\includegraphics[angle=0,scale=0.32]{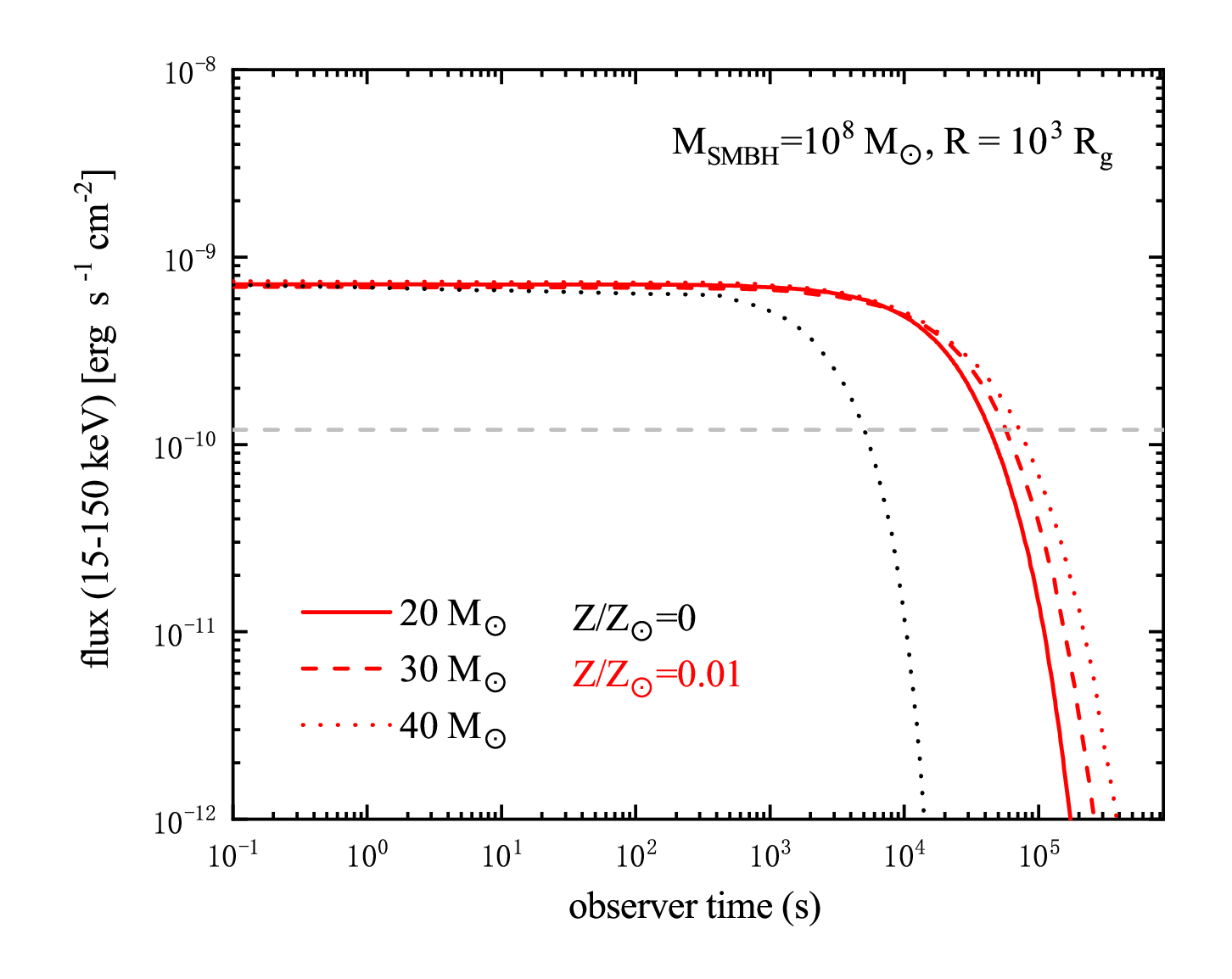}
\caption{Same as Figure \ref{fig5}, but the SMBH mass is $10^{8} M_{\odot}$ and the radial locations from the SMBH is $10^{3} R_{g}$.}
\label{fig7}
\end{figure}

First, we consider the case of the $E_{\rm p}-E_{\rm \gamma,iso}$ correlation with $E_{\rm p} \sim 100$ keV. Figure \ref{fig5} shows the prompt emission light curves of AGN GRBs. The horizontal axis represents the time since jet breakout. The black, red, and blue curves correspond to the cases of progenitors with metallicity of $Z/Z_{\odot}=0$, $0.01$, and $1$, respectively. The solid, dashed, and dotted lines denote progenitor masses of $M_{\rm{pro}}/M_{\odot}$=20, 30, and 40, respectively. The SMBH mass is fixed at $10^{6} M_{\odot}$. The gray dotted line shows the BAT sensitivities $f_{\rm sen} (\Delta t_{\rm obs})$ for an integration time of $\Delta t_{\rm obs}=10^{4}$ s. If the signal flux is larger than $f_{\rm sen} (\Delta t_{\rm obs})$, it can be detected by the BAT within $t_{\rm obs}$. Figures \ref{fig6} and \ref{fig7} present the results for the SMBH masses of $10^{7} M_{\odot}$ and $10^{8} M_{\odot}$, respectively. BAT is expected to detect an AGN GRB as a long-duration X-ray-rich GRB with nearly constant luminosity. The luminosity of the prompt emission mainly depends on the mass accretion rate after the jet breakout. Therefore, the influence of progenitor properties on the prompt emission luminosity is analogous to their influence on the mass accretion rate. For progenitors with zero metallicity, the prompt emission luminosity is the strongest, which can be attributed to the highest late-time mass accretion rate of the central engine. By contrast, for progenitors with solar metallicity, even if the progenitor star is located close to the SMBH, the radiation produced by the jet is still difficult for BAT to detect. In other words, the detection of the prompt emission of AGN GRBs can provide constraints on the density profiles of progenitor stars in AGN disks.

\begin{figure}
\centering
\includegraphics[angle=0,scale=0.32]{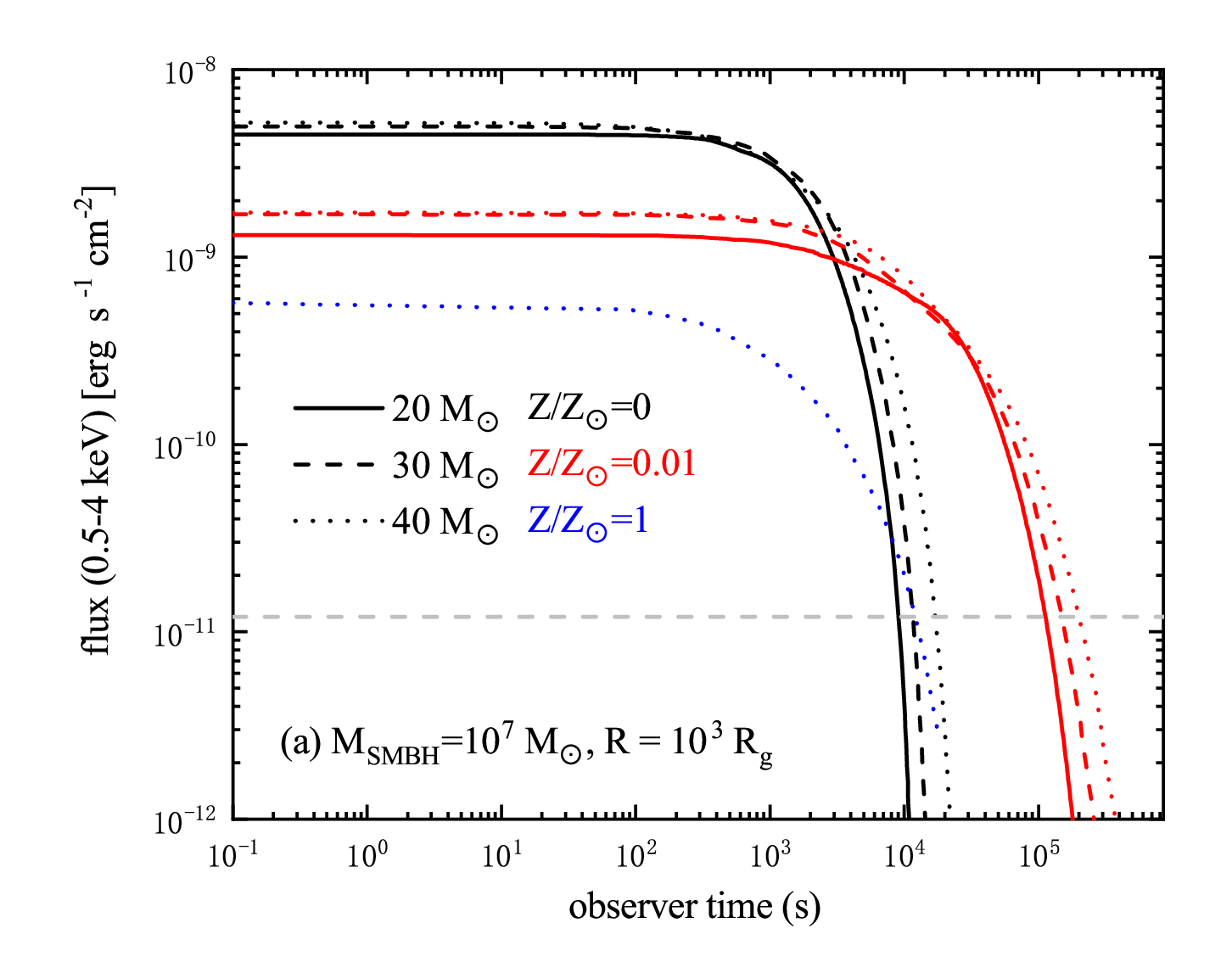}
\includegraphics[angle=0,scale=0.32]{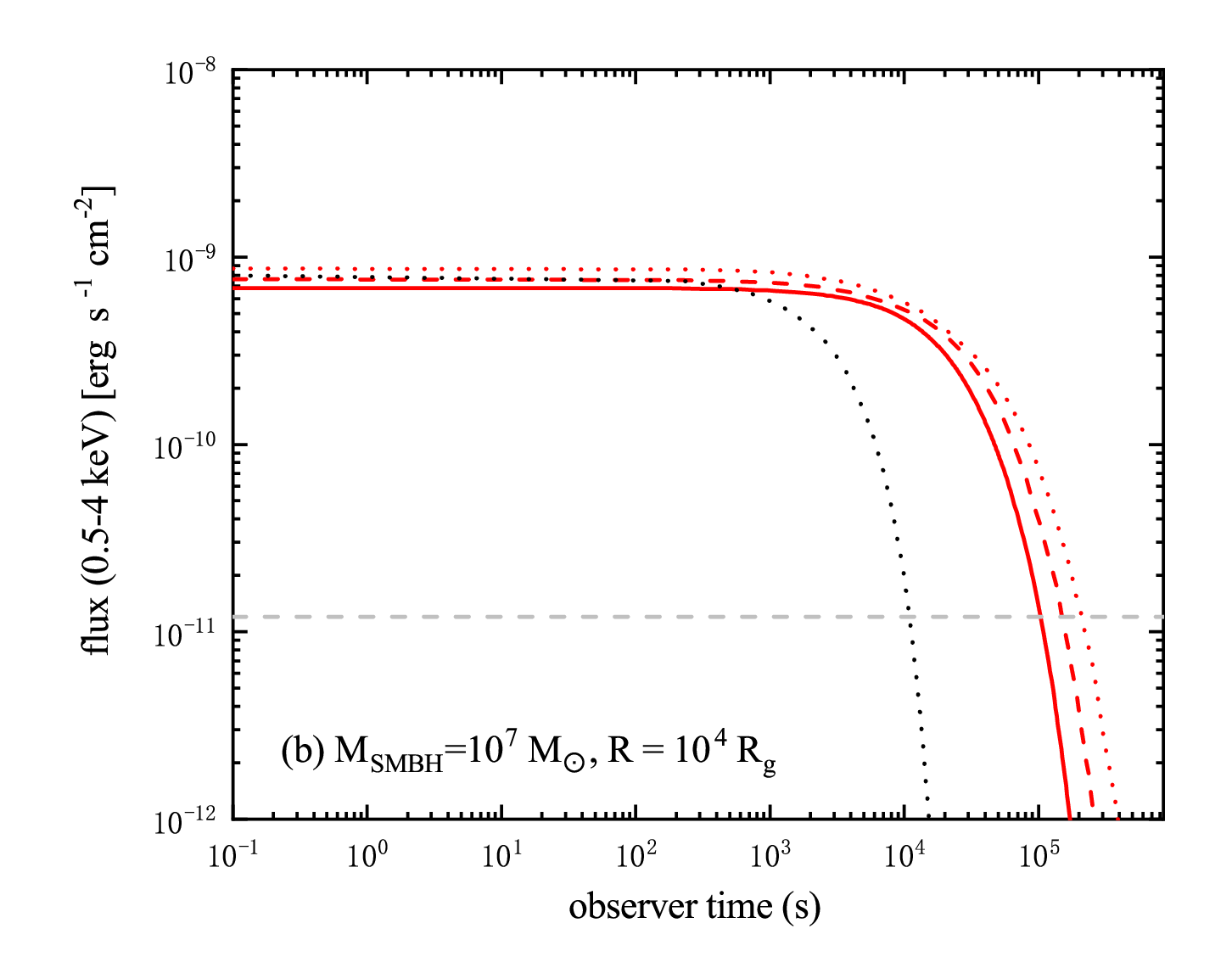}
\caption{Same as Figure \ref{fig5} but for the EP case. The SMBH mass is $10^{7} M_{\odot}$. In panels (a) and (b), the radial locations from the SMBH are $10^{3} R_{g}$ and $10^{4} R_{g}$, respectively.}
\label{fig8}
\end{figure}

We then discuss the detectability of AGN GRBs under the $E_{\rm p}-L_{\rm p}$ correlation. This correlation leads to $E_{\rm p} \sim 5$ keV, which falls within the detectable energy band of EP. The WXT employs Lobster-eye micropore optics, providing an instantaneous field of view of $\sim3600$ square degrees and has a sensitivity of $ \sim 2.6\times 10^{-11} \rm erg\, \rm s^{-1} \rm cm^{-2}$ with a $1000$ s exposure. Figure \ref{fig8} shows the light curves of the prompt emission in the case of the $E_{\rm p}-L_{\rm p}$ correlation. The gray dotted lines indicate the WXT sensitivities $f_{\rm sen} (\Delta t_{\rm obs})$ for an integration time of $\Delta t_{\rm obs}=10^{3}$ s.  One can see that WXT will detect an AGN GRB as a long-duration X-ray flash with nearly constant luminosity.

\subsection{Afterglows}

In this section, we calculate the afterglow emission of AGN GRBs based on the standard external shock model \cite{Toma2011}. The afterglow arises from the external shock driven by the jet in the medium above the AGN disk. In the shocked region, the external shock amplifies the magnetic field and accelerates the electrons to a power-law energy distribution of the form $ N(\gamma _{e})\propto \gamma _{e}^{-p}$. The accelerated electrons produce afterglow emission through synchrotron and synchrotron-self-Compton processes. To calculate the afterglow emission, we assume the number density of the medium outside the AGN disk is comparable to that of the ISM, and set $n=1~\rm cm^{-3}$. We adopt a power-law index $p=2.3$, and set $\epsilon_{B}=0.01$ and $\epsilon_{e}=0.1$ as the fractions of the thermal energy carried by the magnetic field and electrons. Assuming that $10\%$ of the jet energy is converted into the energy for the prompt photon emission, we take the remaining $90\%$ of the jet energy to be available for afterglow emission. For more details on the parameter values, see \citet{Toma2011}. Within this model, we compute the afterglow light curves in the X-ray band ($0.3-10$ keV).

\begin{figure}
\centering
\includegraphics[angle=0,scale=0.32]{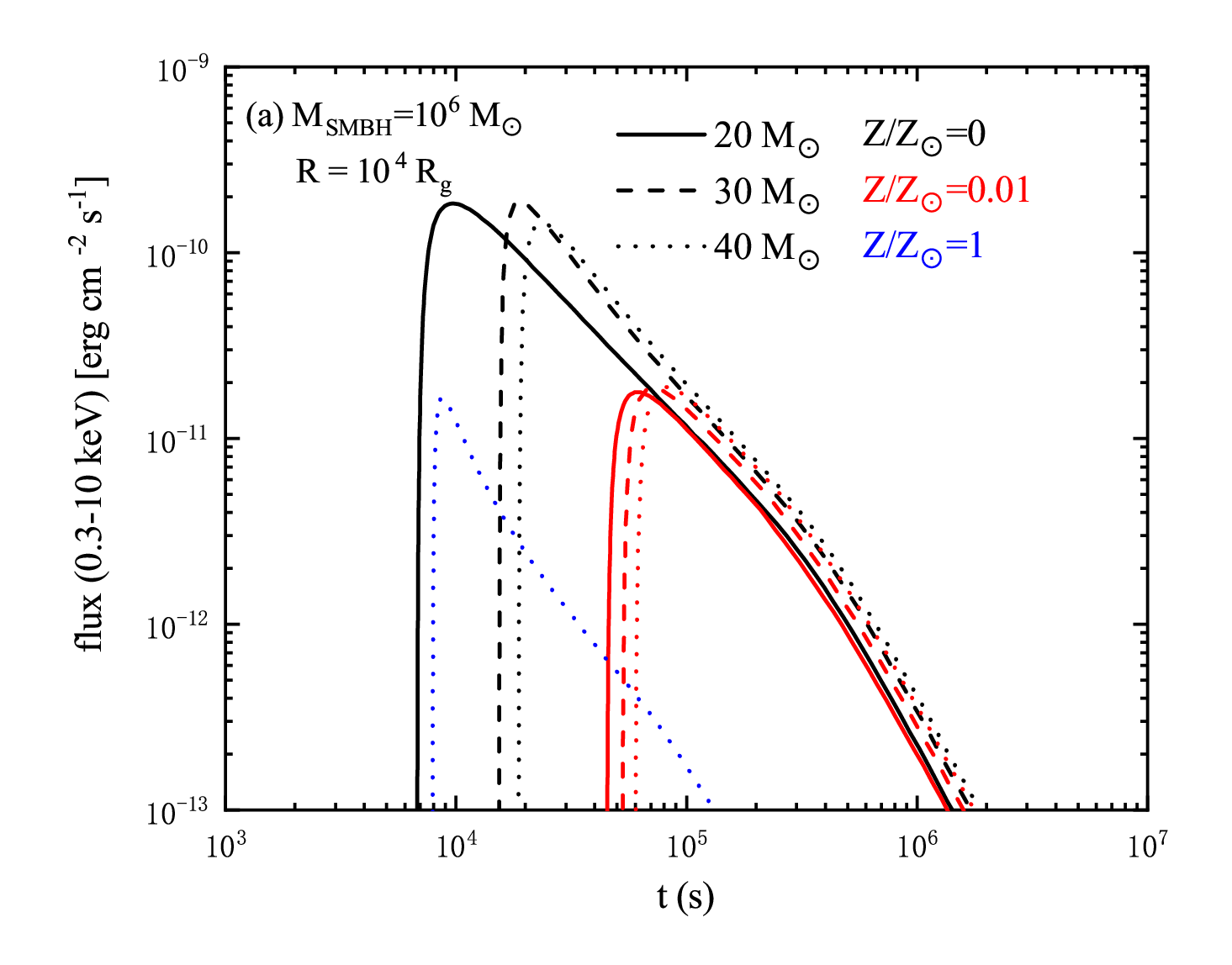}
\includegraphics[angle=0,scale=0.32]{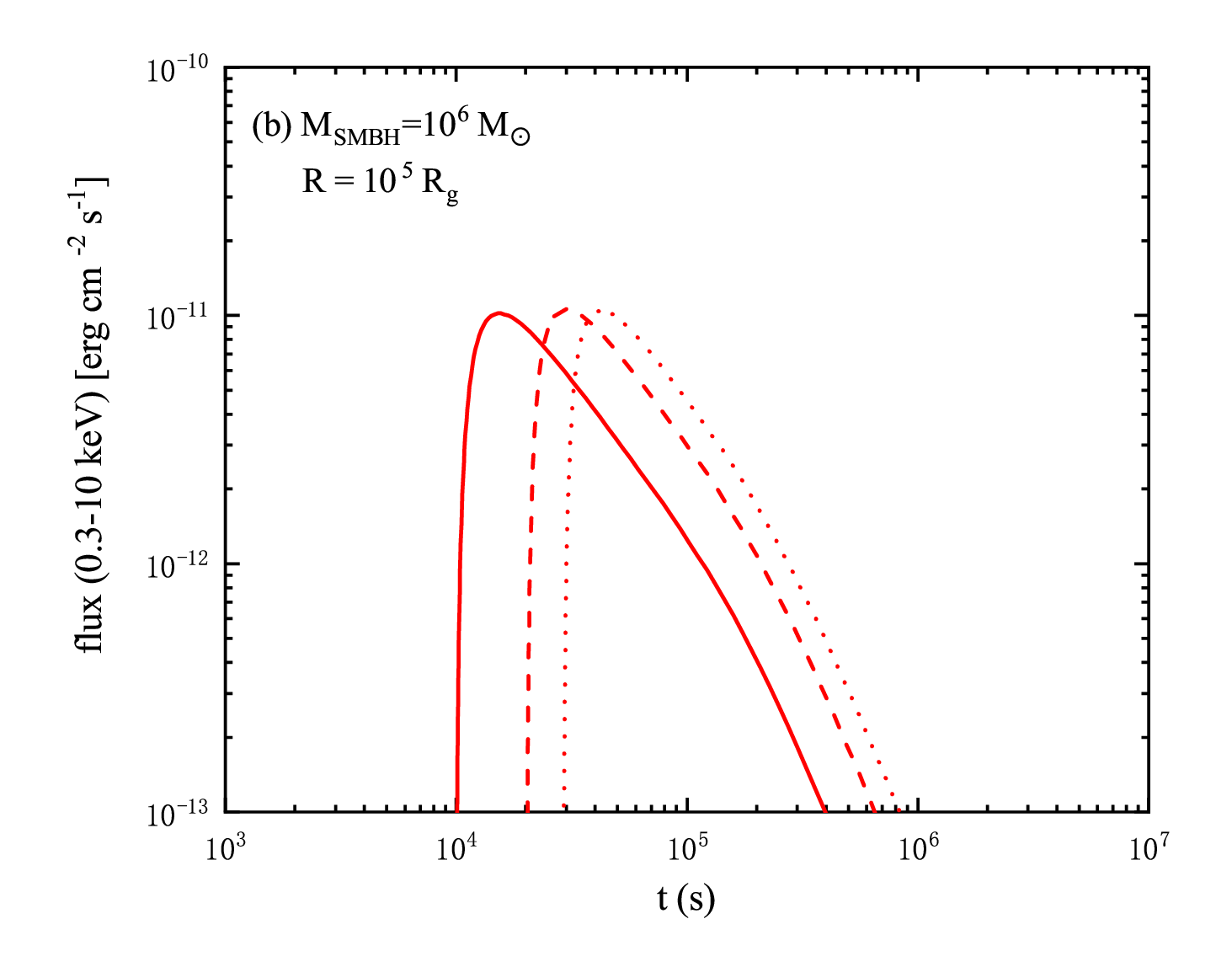}
\caption{X-ray afterglow light curves of AGN GRBs with different progenitors. The black, red, and blue curves correspond to the cases of progenitors with metallicities of $Z/Z_{\odot}=0$, $0.01$, and $1$, respectively. The solid, dashed, and dotted lines represent the progenitor masses of $M_{\rm{pro}}/M_{\odot}=20$, 30, and 40, respectively. The SMBH mass is $10^{6} M_{\odot}$. In panels (a) and (b), the radial locations from the SMBH are $10^{4} R_{g}$ and $10^{5} R_{g}$, respectively.}
\label{fig9}
\end{figure}

\begin{figure}
\centering
\includegraphics[angle=0,scale=0.32]{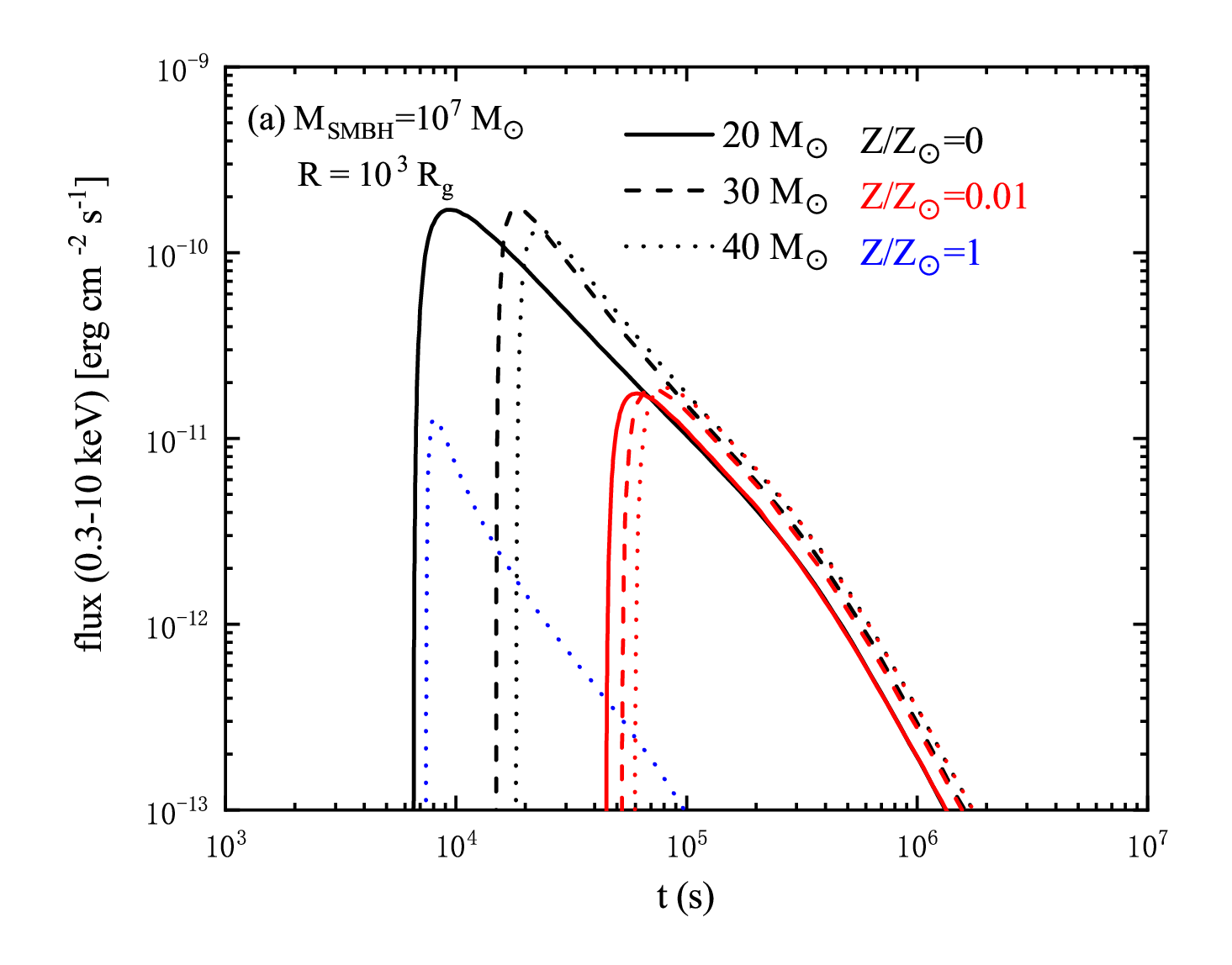}
\includegraphics[angle=0,scale=0.32]{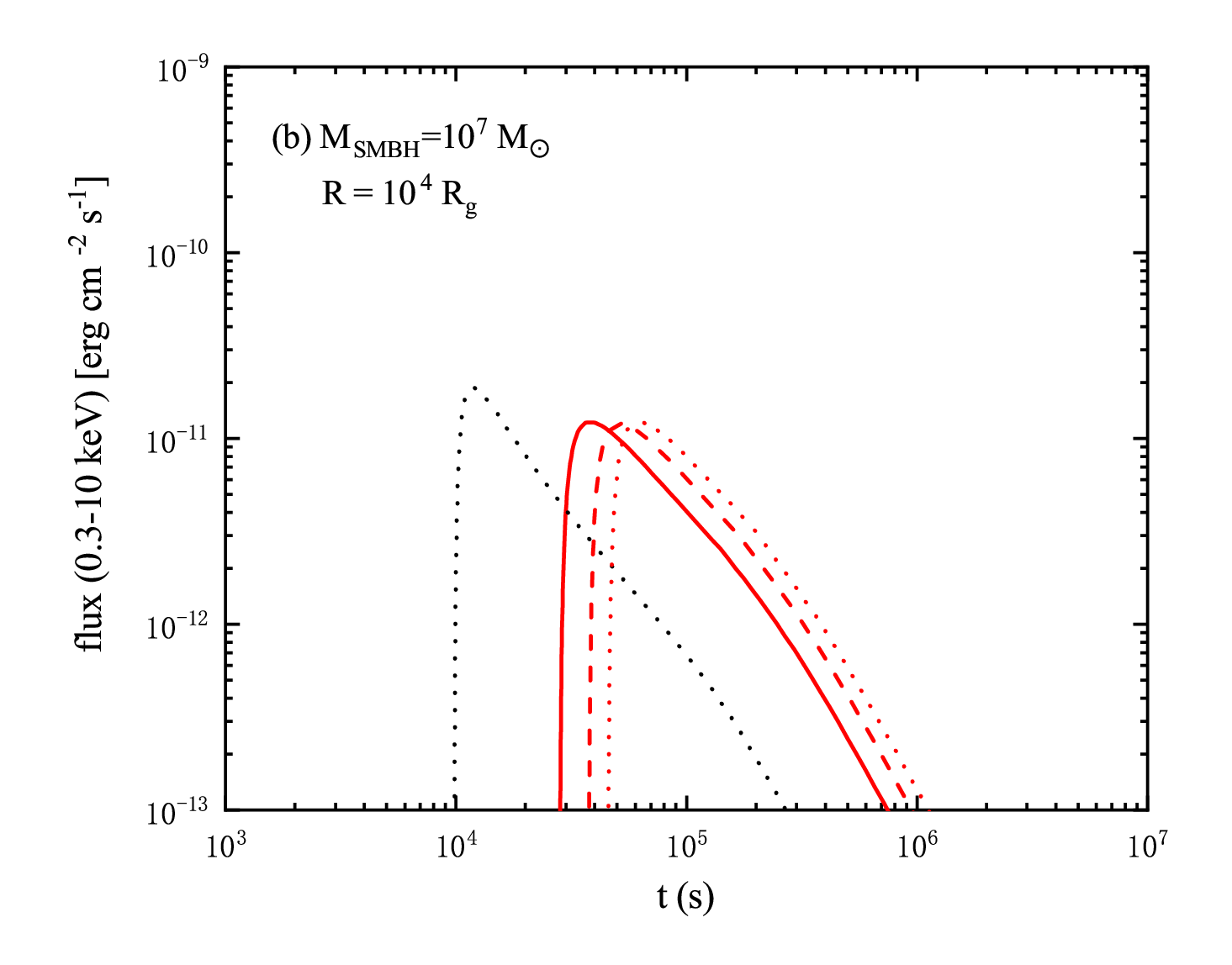}
\caption{Same as Figure \ref{fig9}, but the SMBH mass is $10^{7} M_{\odot}$. In panels (a) and (b), the radial locations from the SMBH are $10^{3} R_{g}$ and $10^{4} R_{g}$, respectively.}
\label{fig10}
\end{figure}

\begin{figure}
\centering
\includegraphics[angle=0,scale=0.32]{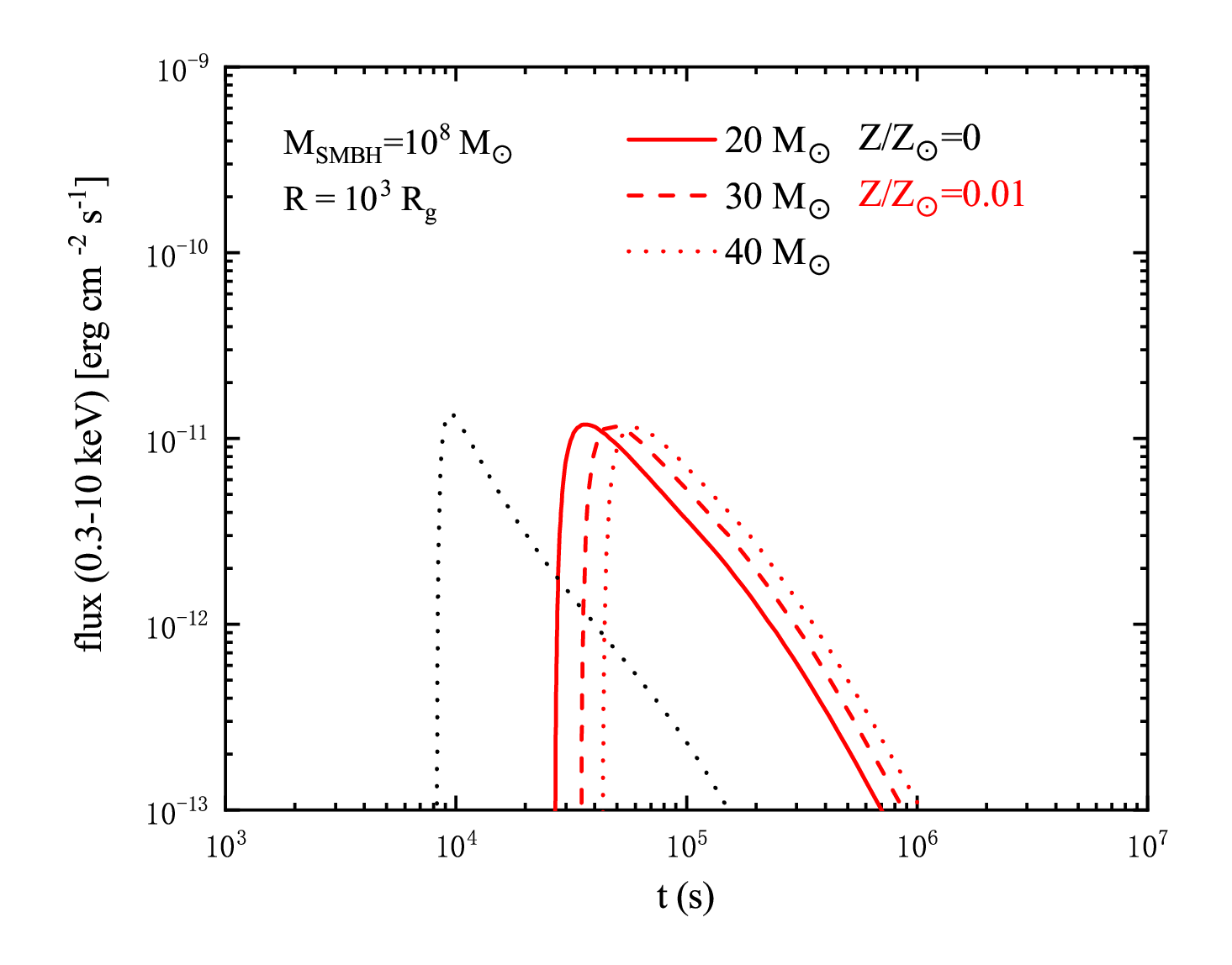}
\caption{Same as Figure \ref{fig9}, but the SMBH mass is $10^{8} M_{\odot}$. In panel(a), the radial locations from the SMBH is $10^{3} R_{g}$.}
\label{fig11}
\end{figure}

Figures \ref{fig9}, \ref{fig10}, and \ref{fig11} display the X-ray afterglow light curves of AGN GRBs. The abscissa represents the time since the beginning of the prompt emission. The results show that the X-ray afterglow emission is detectable by EP FXT at $z=1$, since the sensitivity for FXT can reach the order of $10^{-14}\,\rm erg\,s^{-1}\,cm^{-2}$ with an exposure time of 10 ks. Similar to the prompt emission, the afterglow emission is influenced by progenitor mass and metallicity. Specifically, the external shock emission depends on the isotropic jet energy $E_{\rm iso}$ and the jet duration after breakout, both of which are determined by the progenitor's density profiles. For solar metallicity progenitors, the jet struggles to break out of the disk and the afterglow emission is dim. In contrast, for $Z/Z_{\odot}=0$, the afterglow emission is relatively brighter because of the higher jet luminosity at breakout. Therefore, joint observations of prompt emission and afterglows can provide valuable constraints on the physical properties of progenitor stars embedded in AGN disks.

\section{Conclusions and Discussion}

Massive collapsars are expected to occur within AGN disks. In this work, we investigated jet propagation in the AGN disk and calculated the resulting GRB emission. The effects of progenitor mass and metallicity on GRB emission in AGN disks are studied. In general, ordinary LGRB jets have difficulty breaking out of AGN disks when the mass of the SMBH is large and the explosion site is far away from the SMBH. However, the duration and energy of GRBs strongly depend on the progenitor in the collapsar scenario. The progenitor mass and metallicity can influence the density profiles of stars and, consequently, the duration of the accretion timescale. Low-metallicity progenitor stars experience little mass loss and retain an extended envelope. The prolonged accretion of the massive envelope can significantly extend the activity duration of the central engine, thereby enabling the jet to break out of the AGN disk. 

Here we adopt the SG model to describe AGN disks. Actually, the disk model proposed by \citet{Thompson2005}, hereafter the TQM model, has also been widely applied. \citet{Zhang2024} studied the jet propagation in AGN disks by adopting the SG and TQM models, respectively. The influence of the disk model on jet propagation is complex. In general, the specific criterion for the jet breakout is highly dependent on the AGN disk scale height. For a given central SMBH, the SG model predicts a smaller disk scale height than the TQM model in the inner region of the AGN disk, whereas in the outer region, the trend is reversed. Consequently, in the inner region of the AGN disk, jets are more likely to break out of the SG disk, whereas in the outer region, they can more easily propagate through the TQM disk. A comprehensive understanding of AGN disks is essential for studying jet propagation within them.

Then we evaluate the observational characteristics of AGN GRBs. In this work, we only focus on the GRB emission of successful jets. Their prompt emission is less luminous than that of local LGRBs because of the much longer jet propagation time. Assuming that the $E_{\rm p}-E_{\rm \gamma,iso}$ correlation holds for AGN GRBs, those at $z=1$ may be observed as long-duration X-ray-rich GRBs by the BAT. If the $E_{\rm p}-L_{\rm p}$ correlation holds instead, AGN GRBs may be observed as long-duration X-ray flashes by EP. The X-ray afterglow emission of AGN GRBs at $z=1$ may be detected by EP FXT. The effects of progenitor mass and metallicity on the afterglow emission are analogous to their influence on the prompt emission. Therefore, joint observations of both prompt and afterglow emissions can provide valuable constraints on the density profiles of the progenitor stars in AGN disks. Furthermore, a significant fraction of jets are expected to be choked within the AGN disk. It would be worthwhile to explore how the progenitor properties influence the emission from these choked jets in future work.

In our above calculations, we adopt the pre-SN model as progenitor model. However, due to the dense gas environment, the evolution of these embedded stars can differ significantly from that of stars in the interstellar medium. The mass and spin of these stars undergo rapid evolution through accretion \citep[e.g.,][]{Cantiello2021,Jermyn2021,Jermyn2022}. The fate of stars in AGN disks depends sensitively on their distance from the SMBH and the SMBH mass \citep[e.g.,][]{Dittmann2021,AliDib2023,Fabj2025}. Our work shows that jet propagation in AGN disks is largely determined by the progenitor's density structure, and that envelope accretion facilitates the jet breakout. Moreover,the properties of the progenitor stars might determined by the AGN disk and its environment, so detections of AGN GRBs could give information on the evolution history of AGN disks and therein massive stars.

\section*{acknowledgments}
We thank Prof. Alexander Heger for providing us pre-SN data. This work was supported by the National Key R\&D Program of China (Grant No. 2023YFA1607902), the National Natural Science Foundation of China (Grant Nos. 12173031, 12494572, 12221003, 12303049, and 12503052), the Fund of National Key Laboratory of Plasma Physics (Grant No. 6142A04240201), and the China Manned Space Program (Grant No. CMS-CSST-2025-A13).

\end{document}